\documentclass{article}
\usepackage{arxiv}
\usepackage[utf8]{inputenc} 
\usepackage[T1]{fontenc}    
\usepackage{hyperref}       
\usepackage{url}            
\usepackage{booktabs}       
\usepackage{amsfonts, amsmath}       
\usepackage{nicefrac}       
\usepackage{microtype}      
\usepackage{lipsum}
\usepackage{graphicx}
\graphicspath{ {./images/} }
\usepackage{tabularx}

\title{\LARGE \bf LLM-Mediated Human–AI Interaction in Search and Rescue: Impact of Expertise on Attentional Allocation}

\author{Elahe Oveisi and Hemanth Manjunatha
\thanks{All Authors are with Department of Mechanical and Aerospace Engineering, Oklahoma State University, Stillwater, OK USA 74078-5061}
\thanks{$*$Corresponding Author, 
{\tt\small hemanth.manjunatha@okstate.edu}}
}

\begin{document}
\maketitle
\begin{abstract}
Human-AI teaming (HAT) increasingly involves AI systems that provide real-time, context-aware guidance in complex tasks. While such systems can improve performance, their effectiveness depends on how they shape human cognition and behavior. In particular, AI assistance can introduce cognitive demands and influence attention, planning, and interaction with the task environment, with effects that can vary across levels of expertise. This work investigates these mechanisms in a simulated search and rescue (SAR) environment. We compare human performance under two LLM (Large Language Model)-guided conditions and a no-LLM baseline, and analyze interaction at multiple levels, including task performance, eye-tracking measures, and planning behavior. Eye tracking provides fine-grained insight into attention allocation and interaction with AI guidance, while behavioral measures capture how users structure and adapt their decisions over time. Results indicate that LLM guidance enhanced task efficiency (higher rewards and victims-per-step) but did not increase total victims saved. Eye-tracking data revealed an attention-guidance trade-off, with visual resources shifting to the chat interface alongside increased pupil size variability. Expertise moderated this effect: novices exhibited passive AI reliance, whereas experts maintained a ``verification loop" through persistent environmental scanning. These findings suggest that LLM-mediated teaming efficacy depends on the operator’s ability to cross-reference AI guidance with ground truth to maintain situational awareness.
\footnote{**Accepted for publication at IEEE SMC 2026. © 2026 IEEE.  © 2026 IEEE. Personal use of this material is permitted. Permission from IEEE must be obtained for all other uses, in any current or future media, including reprinting/republishing this material for advertising or promotional purposes, creating new collective works, for resale or redistribution to servers or lists, or reuse of any copyrighted component of this work in other works**}
\end{abstract}

\section{INTRODUCTION}
\begin{figure}[ht!]
    \centering
    \includegraphics[width=1\linewidth]{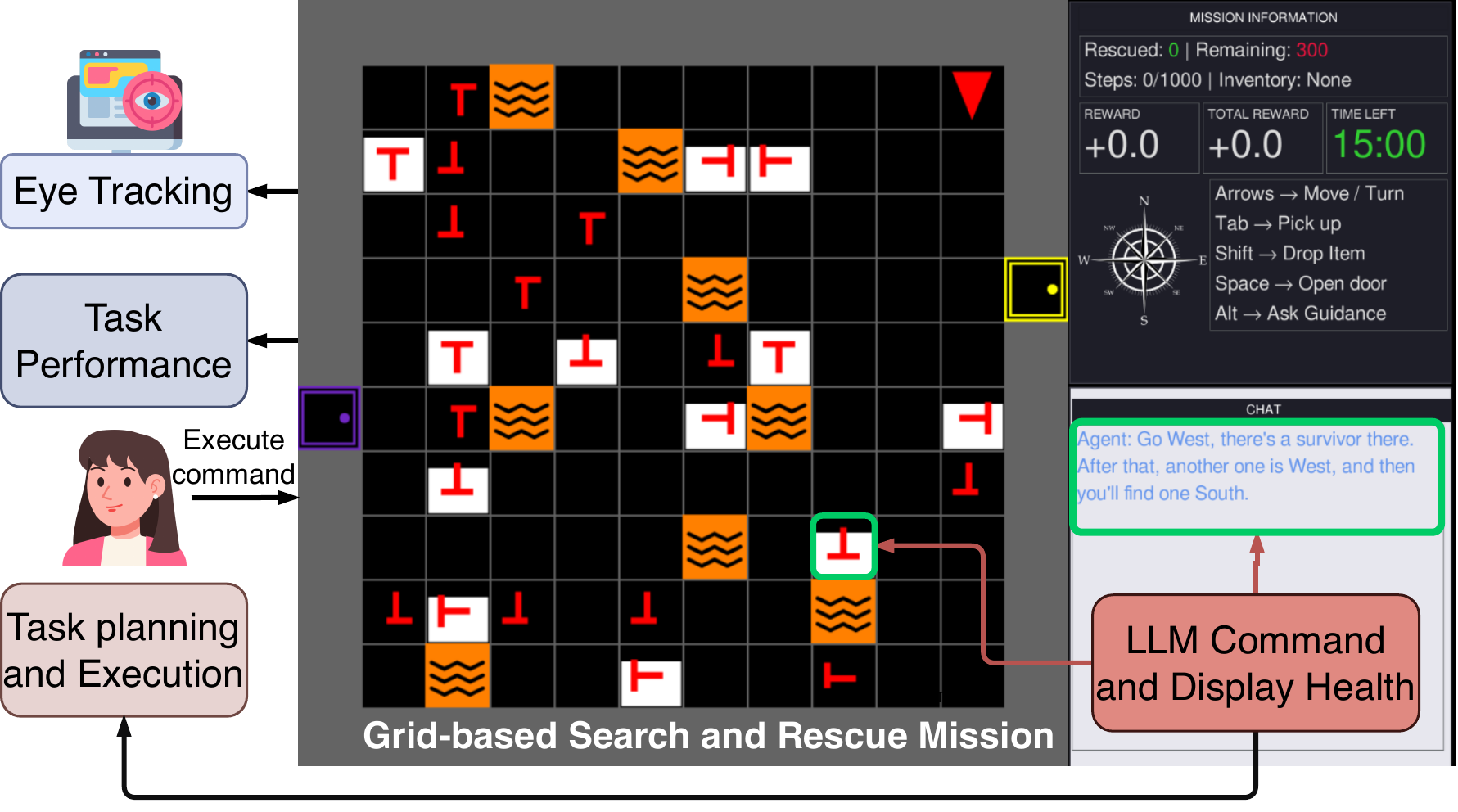}
    \caption{Overview of the experimental framework. Participants perform a grid-based SAR task under different guidance conditions. The study integrates behavioral metrics, eye-tracking measures, and planning-related analysis to examine how LLM-based assistance shapes attention, decision-making, and performance.}
    \label{fig:overview}
\end{figure}

Human-AI teaming (HAT) is emerging as a central paradigm as artificial intelligence systems transition from passive tools to active collaborators in complex tasks~\cite{schmutz2024ai}. While modern AI systems can provide context-aware guidance and natural language recommendations, interacting with such systems introduces additional cognitive demands, requiring users to interpret, evaluate, and integrate AI outputs into their decision-making ~\cite{schmutz2024ai}. As a result, AI assistance may not uniformly improve performance, but instead reshape how humans allocate attention and construct plans, with effects that vary across levels of user expertise. Understanding these interaction mechanisms is therefore critical for designing effective human-AI teams. This need is particularly evident in high-stakes, dynamic domains, such as search and rescue (SAR), where decisions must be made under uncertainty and time pressure. In this work, we examine how AI guidance shapes attention, planning, and performance in the grid-based SAR environment (Fig.~\ref{fig:overview}) through behavioral and eye-tracking analyses.

Search and rescue (SAR) missions are highly cognitively demanding, requiring rapid decision-making under uncertainty, time pressure, and operational risk~\cite{lyu2023unmanned}. Operators must search for targets, interpret incomplete information, and adapt plans in dynamic environments, making SAR a compelling domain for studying human-AI teaming.

Recent advances in large language models (LLMs) provide a particularly relevant instantiation of such guidance. LLMs generate real-time, context-sensitive recommendations through natural language, enabling flexible integration into ongoing decision-making~\cite{handler2024large}. Unlike traditional automation, which delivers fixed alerts or structured outputs, this form of guidance can directly influence how operators interpret the environment, allocate attention, and construct plans during task execution~\cite{handler2024large}.

In high-stakes settings, the effectiveness of AI support depends not only on outcomes but also on how it shapes human decision-making. In SAR, operators must manage competing demands such as monitoring the environment, locating targets, avoiding hazards, and planning routes. AI guidance introduces an additional layer of interaction, requiring users to interpret and evaluate its recommendations, which can increase cognitive load when the guidance is difficult to verify or integrate~\cite{gandhi2023can}. 

Most research on AI-assisted decision-making focuses primarily on performance metrics such as accuracy and completion time~\cite{kovari2024ai}, which provide only a partial view of effectiveness in dynamic tasks. Similar outcomes can arise from fundamentally different interaction dynamics, such as increased monitoring effort or disrupted planning. Understanding LLM-based assistance, therefore, requires examining not only performance, but also how interaction with AI reshapes attention, planning, and decision-making~\cite{leigh2008using}.

These effects can vary substantially across individuals, reflecting differences in expertise, strategy, and cognitive effort that are not captured by performance alone. Capturing such variation requires objective measurements of user behavior, particularly through physiological signals. Eye tracking provides a direct means of measuring visual attention during task performance~\cite{s25134207}. 

In human-AI interaction, eye tracking reveals how users allocate attention between the task environment and AI guidance, including patterns of rechecking and attentional switching~\cite{klingbeil2024trust}. Metrics such as fixation duration, dwell time, and gaze transitions provide insight into cognitive effort, information acquisition, and monitoring behavior~\cite{falkowska2025utilization}. For example, frequent gaze shifts may indicate effort spent verifying AI recommendations, while longer fixations can reflect deeper processing or uncertainty~\cite{castner2024expert}. Integrating eye-tracking with performance measures enables a more comprehensive understanding of how AI influences user behavior.

The present study addresses this gap by comparing three conditions in a simulated SAR environment: two LLM-guided conditions and a baseline without LLM (No-LLM). This design enables evaluation of both the presence of AI assistance and variation in LLM-generated guidance. Rather than focusing solely on end-state performance, we examine behavioral and cognitive processes underlying human-AI interaction, including task outcomes, eye-tracking measures, and planning-related behavior.

By integrating these perspectives, we investigate how LLM guidance influences attention, planning, and performance, and whether different forms of guidance lead to distinct interaction patterns. Importantly, similar performance outcomes may mask differences in cognitive effort and strategy across users with varying levels of expertise. This work therefore shifts the focus from AI outputs to the human process of working with AI, providing insights into how LLM-based guidance shapes behavior and informing the design of more effective human-AI systems for complex, time-critical domains such as SAR.

\section{METHOD AND MATERIALS}
A computer-based search-and-rescue (SAR) environment with a graphical user interface (GUI) was developed for the human subject study. The environment was built on the MiniGrid framework, providing a grid-based, partially observable setting with interactive elements (e.g., doors, keys, hazards, and victims). Participants navigated the environment to efficiently explore, avoid hazards, and rescue victims. Eye movements and behavioral interactions were recorded using a Tobii eye-tracking system. The study was approved by the Institutional Review Board (IRB-26-40-STW).

\subsection{Game Procedure}
The primary task was a computer-based search-and-rescue (SAR) mission implemented as a grid-based navigation environment to study human–AI collaboration. Participants first completed a short training session to learn the controls and interaction rules. During training, they operated in a simplified setting with a single visible room, allowing them to interact with key elements (e.g., doors, keys, hazards, and victims) and understand how actions affected the environment.

\begin{figure}[ht!]
    \centering
    \includegraphics[width=1\linewidth]{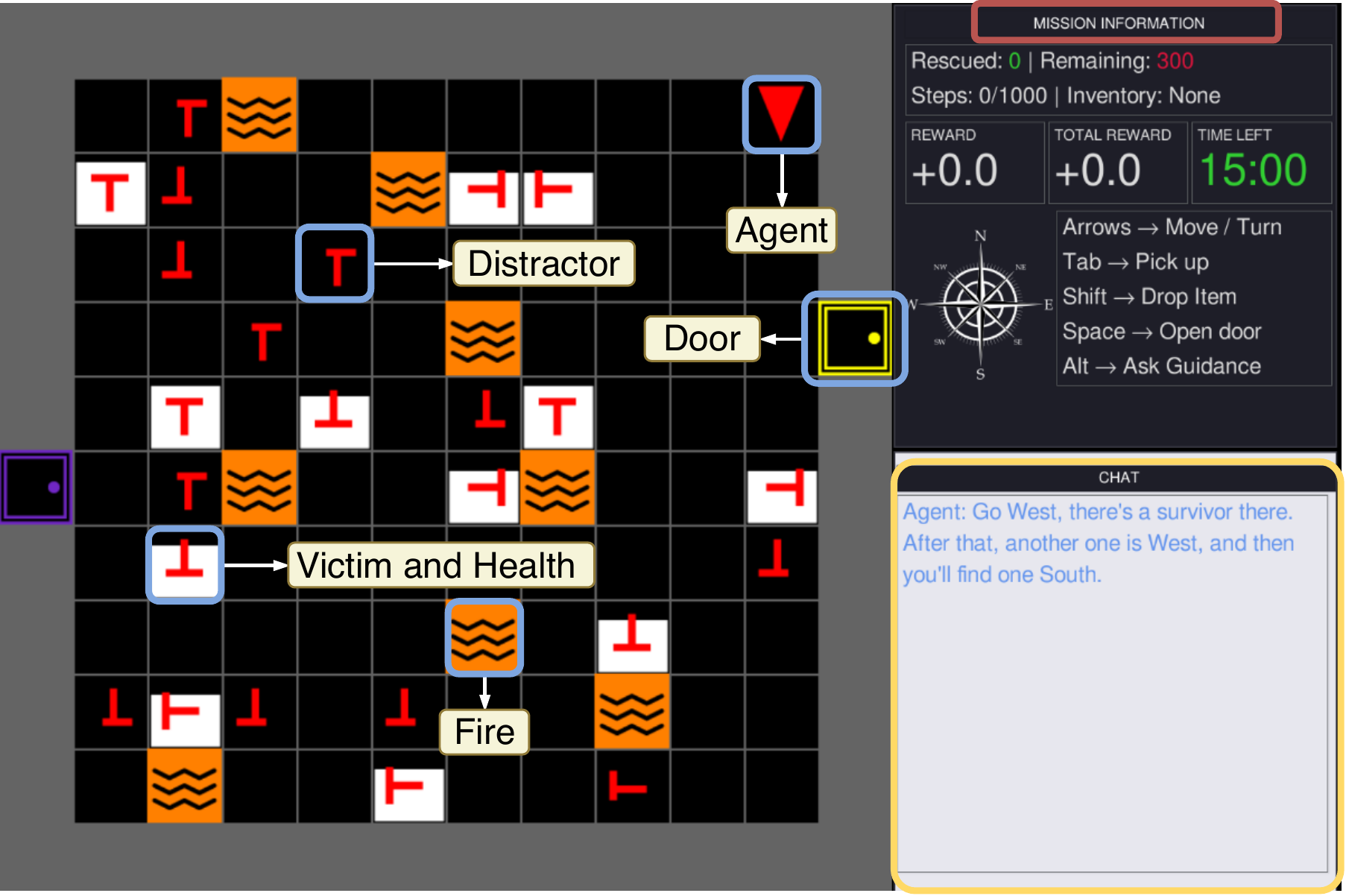}
    \caption{Search-and-rescue (SAR) task environment. Participants navigated a partially observable grid and could receive AI guidance through an in-game interface.}
    \label{fig:game}
\end{figure}

Following training, participants performed the main missions in a partially observable environment, where only nearby cells and objects were visible. Each mission was procedurally generated and consisted of interconnected rooms, corridors, and doors, some of which were locked and required matching colored keys located elsewhere in the environment. The objective was to efficiently explore the environment, avoid hazards (e.g., fire tiles), and rescue victims while minimizing unnecessary movements (Fig.~\ref{fig:game}).

Victims were represented as symmetric “T”-shaped objects, while distractors were slightly asymmetric, requiring visual discrimination. Participants rescued victims by moving adjacent to them and executing a pickup action. All movements, actions, and task events were recorded with timestamps for subsequent analysis.

\begin{figure}[ht!]
    \centering
    \includegraphics[width=1\linewidth]{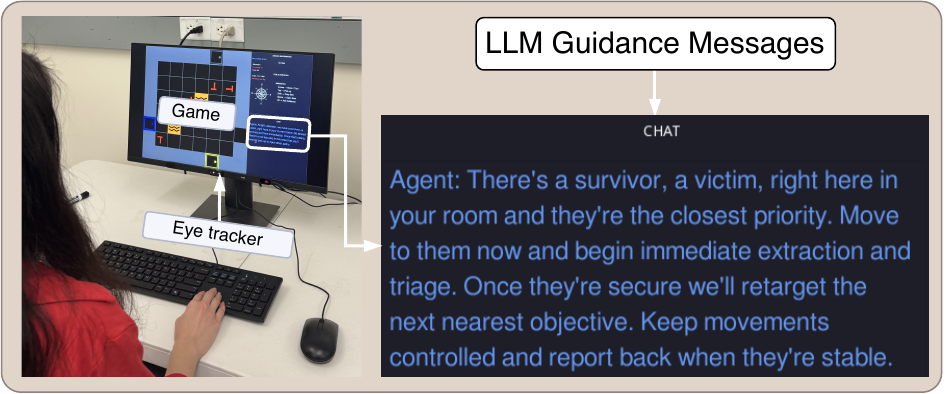}
    \caption{Game interface with AI guidance. Instructions were presented through a chat interface during task execution.}
    \label{fig:participants_llm}
\end{figure}

Each participant completed three 15-minute trials in randomized order: two with LLM assistance and one baseline condition without assistance. Each trial ended either when the time limit was reached or after 1,000 moves, whichever occurred first (Fig~\ref{fig:participants_llm}).

A central feature of the task was progressive victim health deterioration: victims lost health with each step and could die before being reached. Health values were hidden by default and only revealed when participants requested LLM assistance, introducing an information asymmetry in which participants could not continuously monitor victim status, whereas the LLM retained full access. Consequently, participants had to infer deterioration dynamics from environmental structure and indirect cues.

The health deterioration dynamics depended on a victim’s position, facing direction, proximity to hazards such as fire, and surrounding spatial configuration (e.g., doors and open space). Victims near the fire deteriorated more rapidly, particularly when facing toward it, while those farther away declined more slowly. Victims positioned near doors exhibited the lowest fatality rates, whereas downward-facing victims showed consistently higher risk regardless of location. Because these rules were not explicitly disclosed, participants had to learn and reason about risk patterns through repeated observation. Without LLM assistance, this inference process was tedious and error-prone, as it required reconstructing partially hidden dynamics from indirect cues. In contrast, the LLM-assisted condition reduced this burden by revealing victim health and associated cues on demand, effectively functioning as an informational teammate that surfaced otherwise difficult-to-infer dynamics.

In the LLM-assisted conditions, participants could request guidance by pressing \texttt{ALT}. Each request temporarily revealed the victim's health for 5 seconds before being hidden again, along with directional recommendations for movement, routing, and interaction with hazards, keys, and doors. The scoring function further shaped behavior by incentivizing accurate prioritization under uncertainty: rescuing a healthy victim yielded +10 points, selecting a decoy victim incurred a -10 penalty, and rescuing a dead victim incurred a -20 penalty. This asymmetric structure discouraged premature or uninformed actions and reinforced balancing exploratory inference with intermittent LLM-provided information.

\subsection{LLM Guidance Framework}

\begin{figure}[ht!]
    \centering
    \includegraphics[width=\linewidth]{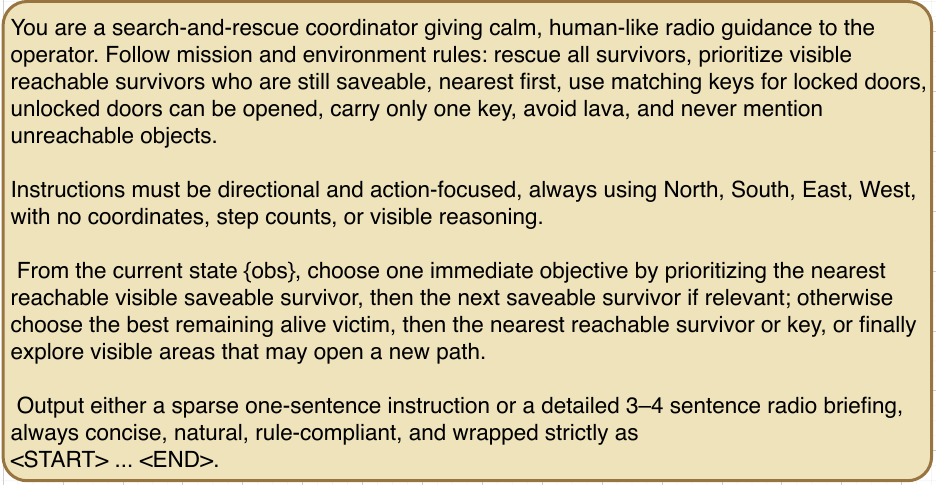}
    \caption{LLM prompt for objective selection and navigation guidance in the SAR task.}
    \label{fig:llm}
\end{figure}

An LLM-based agent was used as a search-and-rescue (SAR) coordinator, providing high-level guidance to the operator during task execution. At each interaction, the model received a structured prompt containing the mission objective, environment rules, and the current observable state in a tabular format (Fig.~\ref{fig:llm}).

To ensure consistent behavior, the prompt enforces a fixed decision hierarchy. The model is instructed to consider only reachable objects and to exclude victims marked as dead. Among reachable and visible survivors, it prioritizes those labeled as savable, i.e., expected to remain alive upon arrival, selecting the nearest based on path length and resolving ties through spatial ordering. If multiple such survivors exist, the model is guided to direct the operator to the nearest first and optionally indicate the next target.

When no savable survivors are available, the prompt directs the model to select the most viable remaining survivor based on projected survivability (e.g., highest margin). If no reachable survivors exist, the objective shifts to the nearest reachable key, door, or unexplored region to enable further progress. The prompt further constrains the model to select only one immediate objective at a time.

Survivability indicators (e.g., savable status and margin) are embedded directly into the state representation, enabling the model to account for time-sensitive constraints during prioritization. In addition, the prompt specifies a strict response format: outputs must be short, action-oriented radio instructions that include directional guidance (North/South/East/West), reference only reachable objects, and avoid low-level details such as coordinates or step counts.

\textbf{Prompt Representation:}
We evaluated three approaches for representing the environment state. An ASCII-based format encoded the grid using symbolic characters, while a semantic format described the scene in natural language (e.g., “A survivor is located to the north behind a locked red door, and a red key lies to the east”). A third approach used a breadth-first search (BFS) to generate a structured, tabular representation capturing key attributes such as object type, visibility, reachability, and path distance from the agent. Among these, the BFS-based structured representation yielded the most consistent and reliable guidance and was therefore used in all experimental trials.

\subsection{Performance Measures}
We evaluate task performance using three complementary metrics capturing efficiency, effectiveness, and overall success in the search-and-rescue task. \textbf{Victims per step} measures efficiency as the number of rescued victims relative to actions taken, reflecting how effectively participants navigated and made decisions. \textbf{Total rewards} captures overall task performance under the scoring function, combining successful rescues with penalties for incorrect or suboptimal actions. \textbf{Number of saved victims} reflects absolute task success, corresponding to the total number of survivors rescued during a trial, independent of efficiency considerations.

\subsection{Eye-Tracking Measures}
Visual attention was measured using a Tobii eye tracker mounted below the monitor, allowing for natural interaction without the need for wearable equipment. The system was calibrated for each participant before the experiment and in-between trials. Eye movements were continuously recorded during both baseline tasks and the main game. The Lab Streaming Layer (LSL) was used to synchronize eye-tracking data with game events.

Several standard eye-tracking metrics were extracted to quantify visual attention. These features capture both fixation-based behavior and broader gaze dynamics. Fixation count refers to the number of times gaze remains relatively stable at a given location, reflecting the allocation of attention \cite{leigh2008using}. Fixation duration captures the length of these stable gaze periods and is associated with visual processing effort \cite{leigh2008using}.

Beyond fixation-based measures, we also extracted complementary oculomotor and perceptual features, including saccade amplitude, fixation areas of interest (AOIs), transitions between AOIs, and the standard deviation of pupil diameter. Together, these metrics provide a more comprehensive characterization of visual search behavior.

Eye movement events were identified using temporal and spatial thresholds adapted to the sampling rate of 60 Hz ($\approx$17.2 ms per sample). Fixations were defined as gaze points lasting at least 100 ms, consistent with prior work \cite{blignaut2009fixation}.

Saccades were detected using duration and velocity constraints appropriate for the sampling rate. An initial minimum event length of 10 ms was used at the detector level \cite{leigh2015neurology}. A velocity threshold of 80 px/s was applied to distinguish saccades from fixation jitter, which refers to small involuntary eye movements during attempted fixation \cite{lappi2016eye}. Saccades longer than 150 ms were excluded as non-physiological or noise-related events \cite{andersson2017one}. 

\subsection{Individual Differences}
\begin{figure}[ht!]
    \centering
    \includegraphics[width=1\linewidth]{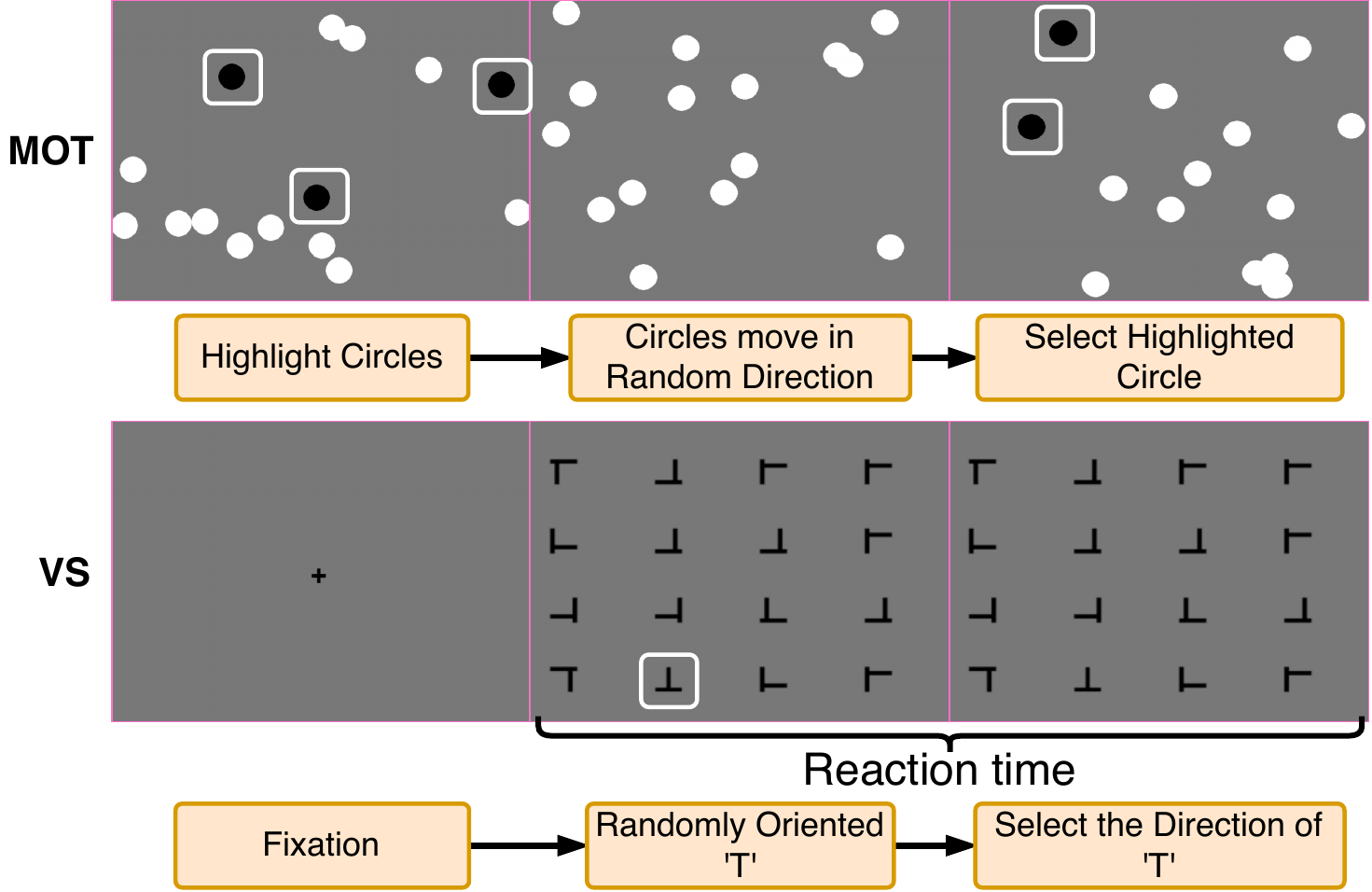}
    \caption{Assessing individual differences in attention and visual search: Multiple Object Tracking (MOT) task measures sustained and divided attention, and the Visual Search (VS) task measures visual discrimination and search efficiency.}
    \label{fig:MOT_VS}
\end{figure}

Before the main task, participants completed two assessment tasks: a Multiple Object Tracking (MOT) and a Visual Search (VS) (Fig.~\ref{fig:MOT_VS}). In the MOT task, participants tracked designated target objects among moving distractors, measuring attentional capacity and dynamic tracking ability \cite{dechterenko2016predicting}. In the VS task, participants identified a target “T” among asymmetric distractors, assessing selective attention and search efficiency \cite{sherman2025designing}. 

Participants were classified into Expert and Novice groups using a K-means clustering algorithm (k=2) to account for baseline differences. This unsupervised approach partitioned individuals based on their performance profiles across three metrics: MOT accuracy, VS accuracy, and VS reaction time, providing a more robust classification than traditional mean-splits.

\subsection{Participants}
Thirteen participants were recruited from the Oklahoma State University population. All participants had normal vision and basic familiarity with computer-based tasks. Informed consent was obtained in accordance with institutional ethical guidelines (IRB-26-40-STW).

\subsection{Data Analysis}
We analyzed run-level (trial-aggregated) outcomes to capture complementary aspects of participant behavior. Run-level analyses provide an overall estimate of the effect of LLM assistance.
Outcomes were analyzed using linear mixed-effects models with a participant-level random intercept. The model was specified as:

\[
\begin{aligned}
y_{ij} =\;& \beta_0 + \beta_1 \text{LLM}_j + \beta_2 \text{Expertise}_i \\
&+ \beta_3 (\text{LLM}_j \times \text{Expertise}_i) + u_i + \varepsilon_{ij}
\end{aligned}
\]

Here, $y_{ij}$ denotes the outcome for participant $i$ in trial $j$. Fixed effects included \textit{LLM condition}, \textit{Expertise}, and their interaction. The term $u_i$ represents a participant-level random intercept to account for repeated observations from the same participant across trials.

Continuous outcomes were modeled directly. Count-based outcomes were transformed using $\log$ before modeling to reduce skewness. Linear mixed-effects models were estimated using restricted maximum likelihood optimization. The Benjamini--Hochberg false discovery rate (FDR) correction was applied across all $p$-values.

\section{Results}
We begin by presenting task performance results, quantified using efficiency (victims per step), total reward, and number of rescued victims, before examining eye-tracking measures of visual attention and cognitive processing under LLM support.

\begin{table}[htbp]
\centering
\footnotesize
\setlength{\tabcolsep}{2pt}
\caption{Run-level linear mixed-effects results for performance outcomes: LLM = main effect of LLM condition (LLM vs.\ no LLM); Exp. = main effect of expertise (expert vs.\ novice); L$\times$E = interaction between LLM condition and expertise; $\beta$ = unstandardized coefficient; SE = standard error; $q$ = Benjamini--Hochberg FDR-corrected $p$-value, ns = not significant}
\label{tab:performance}
\begin{tabular*}{\columnwidth}{@{\extracolsep{\fill}}llrrr}
\toprule
Outcome&Effect&$\beta$&SE&$q$\\
\midrule
\textbf{Victims per step}&\textbf{LLM}&\textbf{0.020}&\textbf{0.006}&\textbf{$<$ 0.01}\\
&Exp.&0.016&0.013&ns\\
&LLM $\times$ Exp.&-0.009&0.010&ns\\
\midrule
\textbf{Total rewards}&\textbf{LLM}&\textbf{267.143}&\textbf{90.203}&\textbf{$<$ 0.01}\\
&Exp.&-17.714&135.183&ns\\
&LLM $\times$ Exp.&181.857&139.742&ns\\
\midrule
\textbf{Saved victims}&LLM&0.035&0.206&ns\\
&Exp.&0.221&0.436&ns\\
&LLM $\times$ Exp.&0.435&0.320&ns\\
\bottomrule
\end{tabular*}
\end{table}

The performance results (Table~\ref{tab:performance}) suggest that LLM support improved task efficiency and overall task quality, rather than simply increasing the number of victims rescued. Participants in the LLM condition achieved significantly higher victims-per-step scores ($\beta = 0.02$, SE $= 0.0063$, $q = 0.0049$) and higher total rewards ($\beta = 267.143$, SE $= 90.2032$, $q = 0.0074$). This indicates that LLM guidance helped participants navigate the search-and-rescue task more efficiently and make better task-related decisions.

However, LLM support did not significantly increase the total number of saved victims ($\beta = 0.0345$, SE $= 0.2064$, $q = 0.8973$). This pattern suggests that the LLM did not simply lead participants to rescue more victims overall; instead, it improved how efficiently they completed the task. Expertise also did not show a significant main effect on performance, and the LLM $\times$ expertise interaction was not significant.

\begin{table}[htbp]
\centering
\footnotesize
\setlength{\tabcolsep}{2pt}
\renewcommand{\arraystretch}{1}
\caption{Run-level linear mixed-effects results for eye-tracking measures associated with information processing. }
\label{tab:eyetracking_info}
\begin{tabular*}{\columnwidth}{@{\extracolsep{\fill}}llrrr}
\toprule
Outcome&Effect&$\beta$&SE&$q$\\
\midrule
\textbf{Mean Fixation Duration}&\textbf{LLM}&\textbf{-19.326}&\textbf{6.216}&\textbf{$<$ 0.01}\\
&Exp.&-26.475&20.594&ns\\
&LLM $\times$ Exp.&17.059&9.629&ns\\
\midrule
\textbf{Pupil Size STD}&\textbf{LLM}&\textbf{0.122}&\textbf{0.014}&\textbf{$<$ 0.0001}\\
&Exp.&-0.042&0.039&ns\\
&\textbf{LLM $\times$ Exp.}&\textbf{-0.051}&\textbf{0.021}&\textbf{$<$ 0.05}\\
\midrule
\textbf{Fixation time in Game Area}&\textbf{LLM}&\textbf{-0.184}&\textbf{0.023}&\textbf{$<$ 0.0001}\\
&Exp.&-0.015&0.037&ns\\
&\textbf{LLM $\times$ Exp.}&\textbf{0.114}&\textbf{0.035}&\textbf{$<$ 0.01}\\
\midrule
\textbf{Fixation time in Chat Area}&\textbf{LLM}&\textbf{0.092}&\textbf{0.016}&\textbf{$<$ 0.0001}\\
&Exp.&0.003&0.022&ns\\
&LLM $\times$ Exp.&-0.045&0.025&ns\\
\midrule
\textbf{Number of Fixations}&LLM&0.248&0.221&ns\\
&Exp.&-0.201&0.305&ns\\
&\textbf{LLM $\times$ Exp.}&\textbf{0.815}&\textbf{0.342}&\textbf{$<$ 0.05}\\
\midrule
\textbf{Number of Saccades}&LLM&0.234&0.210&ns\\
&Exp.&-0.482&0.274&ns\\
&\textbf{LLM $\times$ Exp.}&\textbf{0.961}&\textbf{0.325}&\textbf{$<$ 0.01}\\
\bottomrule
\end{tabular*}
\end{table}

The eye-tracking results (Table~\ref{tab:eyetracking_info}) indicate that LLM support altered both the distribution of visual attention and underlying processing dynamics. Rather than uniformly increasing visual activity, the presence of the LLM primarily redirected attention from direct task monitoring toward the chat interface. This shift was accompanied by reduced mean fixation durations ($\beta = -19.33$, $\mathrm{SE} = 6.22$, $q = 0.005$), suggesting shorter inspection times per visual element and consistent with a partial offloading of information acquisition to the LLM, reducing reliance on sustained visual sampling of the environment.

LLM support was also associated with increased pupil diameter variability ($\beta = 0.12$, $\mathrm{SE} = 0.014$, $q < 0.001$), indicating elevated cognitive load or arousal during integration of external guidance. This effect was attenuated for experts, as shown by the significant LLM $\times$ expertise interaction ($\beta = -0.05$, $\mathrm{SE} = 0.021$, $q = 0.0349$), suggesting reduced sensitivity of experienced participants to the additional cognitive demands introduced by LLM use.

The most pronounced effects were observed in spatial attention allocation. Participants spent less time on the game environment ($\beta = -0.18$, $\mathrm{SE} = 0.023$, $q < 0.001$) and more time on the chat interface ($\beta = 0.09$, $\mathrm{SE} = 0.016$, $q < 0.001$), indicating that the LLM served as a prominent supplementary information source. However, this redistribution was moderated by expertise: experts maintained greater engagement with the environment than novices (LLM $\times$ expertise: $\beta = 0.11$, $\mathrm{SE} = 0.035$, $q = 0.004$) and exhibited stronger LLM-related changes in visual exploration, as reflected in fixation ($\beta = 0.82$) and saccade ($\beta = 0.96$) interaction effects. Together, these results suggest that novices relied more heavily on the LLM for guidance, whereas experts integrated LLM input within a more balanced dual-attention strategy.

\section{Discussion}
The results suggest that LLM support significantly enhanced task efficiency but did not increase overall mission success. Participants achieved higher rewards and more victims rescued per step when the LLM was available, indicating improved navigation and decision efficiency. However, the absence of a significant increase in total rescued victims suggests that the LLM primarily acted as a \textit{local tactical optimizer} rather than expanding overall task capacity.

One interpretation of this pattern is that participants incorporated the LLM as an additional source of task-relevant information rather than relying solely on visual exploration. Consistent with this, eye-tracking results show reduced fixation durations and decreased attention to the game environment, alongside increased attention to the chat interface. This redistribution indicates a trade-off between visual processing of the environment and reliance on external guidance. Increased pupil variability further suggests altered cognitive load during integration of language-based guidance with spatial task demands. Overall, LLM support changed how information was sampled and integrated, rather than simply increasing visual engagement.

Expertise further moderated these effects. Experts maintained greater attention on the game environment and exhibited more active visual scanning, reflected in increased fixation and saccade activity under LLM support. This pattern is consistent with an active verification strategy, in which the LLM serves as a supplementary information source that is cross-checked against the visual environment. In contrast, novices shifted attention more strongly toward the chat interface, indicating greater reliance on LLM guidance. These differences reflect an attention--guidance trade-off in LLM-mediated performance, where external assistance improves efficiency but reshapes visual allocation.

More broadly, these findings provide a baseline for studying trust and error processing in LLM-mediated teaming. Future work will extend this framework to scenarios with incorrect or conflicting AI outputs and incorporate high-density EEG to examine neural correlates of trust calibration and situational awareness, including alpha and theta band dynamics.

\section{Conclusion and Future Work}
This study investigated how LLM-based guidance shapes human-AI teaming in a simulated search-and-rescue (SAR) task. The results reveal an \textit{efficiency–effectiveness gap}: while LLM support significantly improved tactical efficiency, as reflected in higher rewards and more victims rescued per step, it did not increase the total number of victims saved. This suggests that LLM guidance primarily supports local decision-making and navigation efficiency rather than expanding overall task outcomes.

Eye-tracking results further suggest a shift in the operator’s information-processing dynamics. Participants allocated more visual attention to the chat interface and less to the task environment, accompanied by increased pupil size variability, indicating a potential attentional and cognitive trade-off when integrating language-based guidance. Importantly, this trade-off was moderated by expertise. Experts appeared to maintain a ``verification loop," continuing to monitor the environment while using LLM guidance, whereas novices relied more heavily on the AI, potentially reducing situational awareness.

Together, these findings highlight that the effectiveness of LLM-mediated human–AI teaming depends not only on the quality of guidance but also on how users allocate attention and maintain task awareness. From a design perspective, this suggests the need for interfaces that better support attention management, for example, by reducing unnecessary attention shifts or adapting the timing and presentation of guidance based on user expertise. As a next step, future work will extend this framework by integrating additional physiological measures, such as electroencephalography (EEG), to further investigate the cognitive mechanisms underlying human–AI interaction and to inform the development of adaptive assistance systems.
\footnote{Accepted for publication at IEEE SMC 2026. © 2026 IEEE}

\bibliographystyle{IEEEtran}
\bibliography{references}

@Article{s25134207,
AUTHOR = {Jin, Kaizhe and Rubio-Solis, Adrian and Naik, Ravi and Leff, Daniel and Kinross, James and Mylonas, George},
TITLE = {Human-Centric Cognitive State Recognition Using Physiological Signals: A Systematic Review of Machine Learning Strategies Across Application Domains},
JOURNAL = {Sensors},
VOLUME = {25},
YEAR = {2025},
NUMBER = {13},
ARTICLE-NUMBER = {4207},
PubMedID = {40648460},
ISSN = {1424-8220},
DOI = {10.3390/s25134207}
}

@article{klingbeil2024trust,
  title={Trust and reliance on AI—An experimental study on the extent and costs of overreliance on AI},
  author={Klingbeil, Artur and Gr{\"u}tzner, Cassandra and Schreck, Philipp},
  journal={Computers in Human Behavior},
  volume={160},
  pages={108352},
  year={2024},
  publisher={Elsevier}
}

@article{lyu2023unmanned,
  title={Unmanned aerial vehicles for search and rescue: A survey},
  author={Lyu, Mingyang and Zhao, Yibo and Huang, Chao and Huang, Hailong},
  journal={Remote Sensing},
  volume={15},
  number={13},
  pages={3266},
  year={2023},
  publisher={MDPI}
}

@book{leigh2008using,
  title={Using Eye Movements as an Experimental Probe of Brain Function: A Symposium in Honor of Jean B{\"u}ttner-Ennever},
  author={Leigh, R John and Kennard, Christopher},
  volume={171},
  year={2008},
  publisher={Elsevier}
}

@article{gandhi2023can,
  title={How can artificial intelligence decrease cognitive and work burden for front line practitioners?},
  author={Gandhi, Tejal K and Classen, David and Sinsky, Christine A and Rhew, David C and Vande Garde, Nikki and Roberts, Andrew and Federico, Frank},
  journal={JAMIA open},
  volume={6},
  number={3},
  pages={ooad079},
  year={2023},
  publisher={Oxford University Press}
}

@article{castner2024expert,
  title={Expert gaze as a usability indicator of medical AI decision support systems: a preliminary study},
  author={Castner, Nora and Arsiwala-Scheppach, Lubaina and Mertens, Sarah and Krois, Joachim and Thaqi, Enkeleda and Kasneci, Enkelejda and Wahl, Siegfried and Schwendicke, Falk},
  journal={NPJ Digital Medicine},
  volume={7},
  number={1},
  pages={199},
  year={2024},
  publisher={Nature Publishing Group UK London}
}

@article{kovari2024ai,
  title={AI for decision support: Balancing accuracy, transparency, and trust across sectors},
  author={Kovari, Attila},
  journal={Information},
  volume={15},
  number={11},
  pages={725},
  year={2024},
  publisher={MDPI}
}

@article{falkowska2025utilization,
  title={Utilization of Eye-Tracking Metrics to Evaluate User Experiences—Technology Description and Preliminary Study},
  author={Falkowska, Julia and Sobecki, Janusz and Falkowski, Micha{\l}},
  journal={Sensors},
  volume={25},
  number={19},
  pages={6101},
  year={2025},
  publisher={MDPI}
}

@inproceedings{dechterenko2016predicting,
  title={Predicting eye movements in multiple object tracking using neural networks},
  author={Dechterenko, Filip and Lukavsky, Jiri},
  booktitle={Proceedings of the Ninth Biennial ACM Symposium on Eye Tracking Research \& Applications},
  pages={271--274},
  year={2016}
}

@article{sherman2025designing,
  title={Designing a test battery for real-world visual search},
  author={Sherman, Charli and Clarke, Alasdair DF and Hughes, Anna E},
  journal={Scientific Reports},
  volume={15},
  number={1},
  pages={39430},
  year={2025},
  publisher={Nature Publishing Group UK London}
}

@article{andersson2017one,
  title={One algorithm to rule them all? An evaluation and discussion of ten eye movement event-detection algorithms},
  author={Andersson, Richard and Larsson, Linnea and Holmqvist, Kenneth and Stridh, Martin and Nystr{\"o}m, Marcus},
  journal={Behavior research methods},
  volume={49},
  number={2},
  pages={616--637},
  year={2017},
  publisher={Springer}
}

@article{lappi2016eye,
  title={Eye movements in the wild: Oculomotor control, gaze behavior \& frames of reference},
  author={Lappi, Otto},
  journal={Neuroscience \& Biobehavioral Reviews},
  volume={69},
  pages={49--68},
  year={2016},
  publisher={Elsevier}
}

@article{blignaut2009fixation,
  title={Fixation identification: The optimum threshold for a dispersion algorithm},
  author={Blignaut, Pieter},
  journal={Attention, Perception, \& Psychophysics},
  volume={71},
  number={4},
  pages={881--895},
  year={2009},
  publisher={Springer}
}

@book{leigh2015neurology,
  title={The neurology of eye movements},
  author={Leigh, R John and Zee, David S},
  year={2015},
  publisher={Oxford university press}
}

@article{schmutz2024ai,
  title={AI-teaming: Redefining collaboration in the digital era},
  author={Schmutz, Jan B and Outland, Neal and Kerstan, Sophie and Georganta, Eleni and Ulfert, Anna-Sophie},
  journal={Current Opinion in Psychology},
  volume={58},
  pages={101837},
  year={2024},
  publisher={Elsevier}
}

@article{handler2024large,
  title={Large language models present new questions for decision support},
  author={Handler, Abram and Larsen, Kai R and Hackathorn, Richard},
  journal={International Journal of Information Management},
  volume={79},
  pages={102811},
  year={2024},
  publisher={Elsevier}
}

\end{document}